\documentclass[%
 reprint, 
 floatfix,
 twocolumn,
 10pt
]{revtex4-2}

\usepackage{amsmath,amssymb}

\usepackage{placeins}
\usepackage{comment}
\usepackage{graphicx}
\usepackage{dcolumn}
\usepackage{bm}

\usepackage{amsmath}
\usepackage{multirow}
\usepackage{wrapfig}
\usepackage{graphicx}
\usepackage{float}
\usepackage{hyperref}
\usepackage{amssymb}
\usepackage{bbold}
\usepackage{color}
\usepackage{ulem}
\usepackage{comment}
\usepackage{quantikz}
\usepackage{physics}
\usepackage{hhline}
\usepackage{natbib}
\usepackage[separate-uncertainty=false]{siunitx}
\usepackage{adjustbox}
\usepackage{array}
\newcolumntype{P}[1]{>{\centering\arraybackslash}p{#1}}

\newcommand{\RX}{\ensuremath{\mathrm{R_x}}\ }
\newcommand{\RY}{\ensuremath{\mathrm{R_y}}\ }
\newcommand{\RZ}{\ensuremath{\mathrm{R_z}}\ }

\begin{document}
\preprint{APS/123-QED}
\title{Benchmarking simulated and physical quantum processing units\\using quantum and hybrid algorithms}

\author{Mohammad~Kordzanganeh}
\author{Markus~Buchberger}
\author{Basil~Kyriacou}
\author{Maxim~Povolotskii}
\author{Wilhelm~Fischer}
\author{Andrii~Kurkin}
\author{Wilfrid~Somogyi}
\author{Asel~Sagingalieva}
\author{Markus~Pflitsch}
\author{Alexey~Melnikov}\thanks{CONTACT Alexey~Melnikov. Email: ame@terraquantum.swiss.
\begin{center}
\fbox{
\begin{minipage}{0.47\textwidth}
Kindly refer to the updated and published version of the paper, replete with the most recent additions and revisions: \mbox{Mohammad} Kordzanganeh, Markus Buchberger, Basil Kyriacou, Maxim \mbox{Povolotskii}, Wilhelm Fischer, Andrii Kurkin, Wilfrid Somogyi, Asel Sagingalieva, Markus Pflitsch, Alexey Melnikov. Benchmarking simulated and physical quantum processing units using quantum and hybrid algorithms. Adv. Quantum Technol. 6, 2300043 (2023), DOI: \href{https://doi.org/10.1002/qute.202300043}{10.1002/qute.202300043}
\end{minipage}
}
\end{center}}

\affiliation{Terra Quantum AG, 9000 St.~Gallen, Switzerland}
\affiliation{QMware AG, 9000 St.~Gallen, Switzerland}


\begin{abstract}
Powerful hardware services and software libraries are vital tools for quickly and affordably designing, testing, and executing quantum algorithms. A robust large-scale study of how the performance of these platforms scales with the number of qubits is key to providing quantum solutions to challenging industry problems. This work benchmarks the runtime and accuracy for a representative sample of specialized high-performance simulated and physical quantum processing units. Results show the QMware simulator can reduce the runtime for executing a quantum circuit by up to 78\% compared to the next fastest option for algorithms with fewer than 27 qubits. The AWS SV1 simulator offers a runtime advantage for larger circuits, up to the maximum 34 qubits available with SV1. Beyond this limit, QMware can execute circuits as large as 40 qubits. Physical quantum devices, such as Rigetti's Aspen-M2, can provide an exponential runtime advantage for circuits with more than 30 qubits. However, the high financial cost of physical quantum processing units presents a serious barrier to practical use. Moreover, only IonQ's Harmony quantum device achieves high fidelity with more than four qubits. This study paves the way to understanding the optimal combination of available software and hardware for executing practical quantum algorithms.
\end{abstract}

\maketitle


\section{Introduction}
\label{sec:introduction}

Quantum computing is a rapidly growing field of technology with increasingly useful applications across both industry and research. This new paradigm of computing has the potential to solve classically-intractable problems, by exploiting an exponentially-increasing computational space. This allows quantum algorithms to dramatically reduce the runtime for solving computationally resource-intensive problems.

There is a plethora of quantum algorithms, of which parameterized quantum circuits represent the most general form. Quantum neural networks (QNNs) are quantum machine learning (QML) algorithms~\cite{biamonte2017quantum,dunjko2018machine,lamata2020quantum,qml_review_2023} that leverage powerful techniques developed for classical neural networks, to optimize this parameterized structure, and have already been applied to solve a number of industrial problems~\citep{alc_cla_2020,sag_hyp_2022,pis_qua_2021,senokosov2023quantum,rud_gen_2022,sag_hyb_2022,rainjonneau2023quantum,sedykh2023quantum}. The complexity and performance of classical neural networks employed to solve data-intensive problems has grown dramatically in the last decade. Although algorithmic efficiency has played a partial role in improving performance, hardware development (including parallelism and increased scale and spending) is the primary driver behind the progress of artificial intelligence~\citep{ber_fre_2021, her_mea_2020}. Unlike their classical counterparts, QNNs are able to learn a generalized model of a dataset from a substantially smaller training set~\citep{per_pra_2022,car_gen_2022,abb_pow_2021} and typically have the potential to do so with polynomially or exponentially simpler models~\citep{boi_cha_2018, ser_equ_2004, ris_dem_2017}. Thus, they provide a promising opportunity to subvert the scaling problem encountered in classical machine learning~\citep{ben_par_2019, sch_sup_2018, alc_cla_2020, kordzanganeh2023parallel, emm_qua_2022, coy_qua_2021, coe_fou_2020, mei_gra_2020}, which presents a serious challenge for data-intensive problems that are increasingly bottle-necked by hardware limitations~\citep{xu_sca_2018, sze_eff_2017, hor_com_2014}.

Nonetheless, even for a small dataset, training QNNs requires on the order of a million circuit evaluations. This is a consequence of the multiplicative number of data points, evaluations required for calculating the gradient~\citep{sch_eva_2019}, and iterations before a solution is reached. This makes them a relatively challenging and resource-intensive use case for quantum processing units (QPU). Therefore, QNNs require stable, on-demand, and accurate quantum circuit execution. A plethora of different options for executing quantum circuits exist. These are either physical QPUs or classical hardware simulating quantum behaviour. In both cases multiple vendor options and services are available. Establishing the combination of software and hardware that provides the optimum runtime, cost, and accuracy is crucial to the future of democratizing quantum software development.

In this study, different pairings of software development kits (SDKs) and hardware platforms are compared in order to determine the fastest and most cost-efficient route to developing novel quantum algorithms. This benchmark is performed using QNNs, which represent the most general form of a quantum algorithm. The benchmark indicates an advantage in using the QMware basiq simulator for circuits with 2 to 26 qubits, AWS SV1 for 28-34 qubits, and QMware basiq for 36-40 qubits. Additionally, QPUs from four different vendors (IonQ, Oxford Quantum Circuits (OQC), IBM, and Rigetti) were benchmarked for runtime, accuracy, and cost. The results show QPUs could become time-competitive in a practical use case for circuits with 30 qubits or more. However, the current low fidelity attained by many of these systems precludes their application to industrial problems. The Python implementation code for the benchmarks presented in this study is available in Ref.~\cite{benchmarking_github}.

This investigation is not an exhaustive benchmark of all available systems, and it is worth acknowledging the existence of other state-of-the-art services. This includes hardware quantum simulators such as IBM's simulator statevector and Atos's Quantum Learning Machine~\citep{ato_qua_2022}, as well as software backends such as the IBM Qiskit machine learning suite and the Qulacs package~\citep{aleksandrowicz2019qiskit,aoyama_2022_qulacs}. It also includes QPUs such as IBM's Eagle processor~\citep{chow_2021_ibm}, Honeywell's System Model H1~\citep{pino_2021_demonstration}, and Google's Sycamore processor~\citep{arute_2019_quantum}. The authors also acknowledge other works that involved a similar technique in performing benchmarks on quantum hardware. These include hardware specific performance benchmarks such as Refs.~\citep{wri_2019_ben,mcc_2019_qua} as well as metric-specific tests such as Refs.~\citep{sal_2019_ben,lub_2021_app,pac_2022_qas,mil_2021_app,a2022_qscore,dal_2020_ana,mesman_2021_qpa}.

The work is organized in five sections. Sec.~\ref{sec:runtime-benchmarking} describes the methodology, including the benchmark algorithms, the hardware and software tested, as well as the results of the runtime benchmark. Sec.~\ref{sec:accuracy-and-cost-benchmarking} details the cost of executing the proposed quantum circuits on various QPUs and compares their fidelity. Sec.~\ref{sec:discussion} discusses the execution of large quantum circuit, with as many as 40 qubits, and the implementation of multi-threading in simulators. Finally Sec.~\ref{sec:summary} provides a summary of the findings, and outlines key challenges for the quantum computing industry.

\section{Runtime Benchmarking}
\label{sec:runtime-benchmarking}

\begin{figure*}[t]
    \centering
    \includegraphics[width=\textwidth]{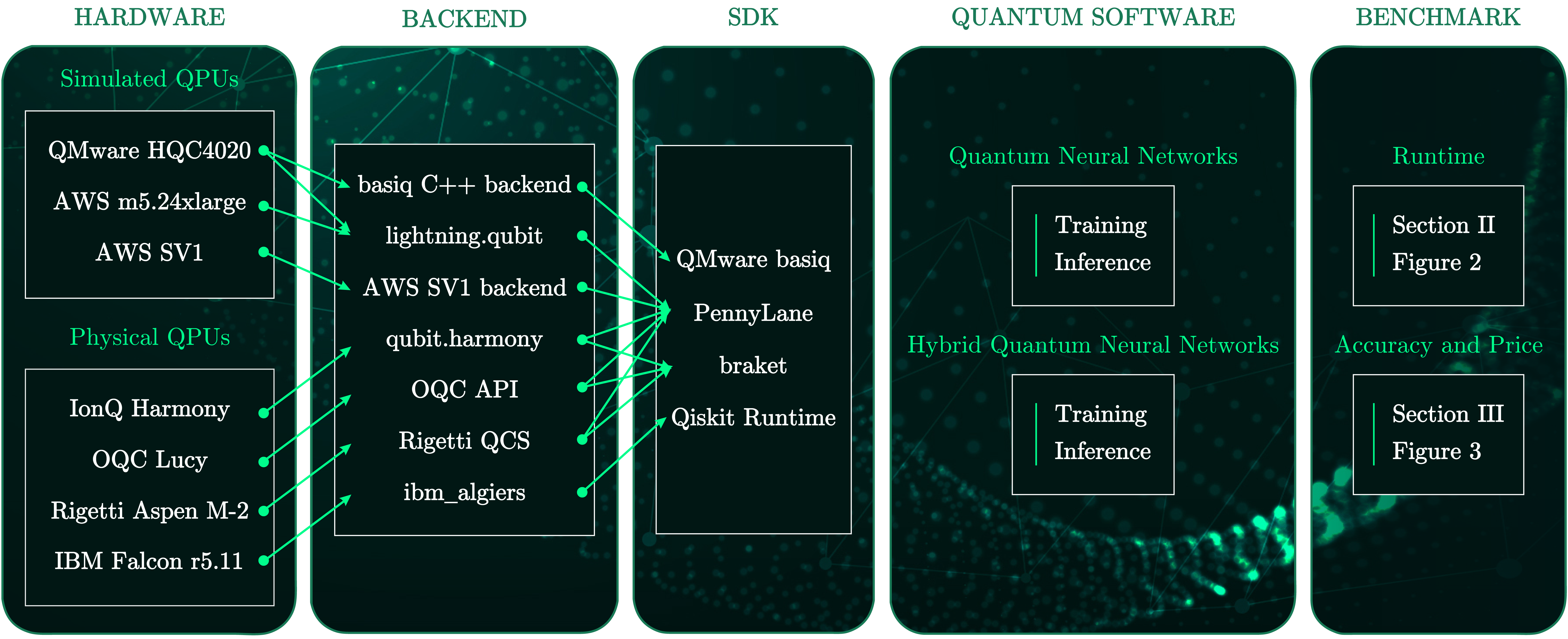}
\caption{The scheme for the benchmark. Simulated and native QPU stacks are benchmarked in the present work using open-source software and proprietary software on AWS, IonQ, OQC, Rigetti, IBM, and QMware.}
    \label{fig:benchmarking-scheme}
\end{figure*}

Quantum simulators are bipartite systems, consisting of a software library and the hardware on which the software is run. Both play a crucial role in the development and execution of a quantum circuit. Although the various software libraries available for quantum simulation are often implemented in a hardware-agnostic way, the \textit{internal} implementation of the linear algebraic methods and the manner in which quantum logic gates are compiled will have a significant impact on performance. This is true for the execution of any quantum circuit, but particularly relevant for gradient calculations when optimizing a QNN during the training phase, due to the use of gradient-based optimization techniques. In particular, the standard \textit{parameter-shift} method of calculating the gradient of the circuit output with respect to each of the $n$ trainable parameters increases the number of expectation values evaluated by a factor of $2n$. By comparison, the forward pass of a trained QNN requires evaluating just a single expectation value, which can usually be obtained with fewer than one thousand circuit shots. The specification of the simulator hardware also plays an important role in the ability to quickly and efficiently optimize a variational quantum algorithms. Moreover, the synergy between software and hardware influences how much computational overhead is required and the ease with which quantum algorithms can be designed, tested, and deployed.

For many quantum computing applications, the open-source PennyLane Python library is extremely popular~\citep{ber_pen_2022}. It is also the recommended quantum simulation library for the AWS Braket computing service, which provides a performance-optimized version of the PennyLane library~\citep{ama_ama_2020}. PennyLane offers a variety of qubit devices. The most commonly used is the \texttt{default.qubit}, which implements a Python backend using NumPy, TensorFlow, JAX, and PyTorch. More recently the \texttt{lightning.qubit} device was introduced, which implements a high-performance C++ backend~\citep{ber_pen_2022}. QMware's cloud computing service provides a quantum simulator stack which supports open-source, hardware-agnostic libraries, such as PennyLane, in addition to QMware's bespoke quantum computing Python library \textit{basiq}~\citep{qmw_qmw_2022}. The basiq library also supports a PennyLane plugin, which translates circuits built using the Pennylane SDK into circuits that can be executed using the basiq backend.

\subsection{Methodology}
\label{sec:methodology}

\begin{figure*}[th!]
    \centering
    \includegraphics[width=\textwidth]{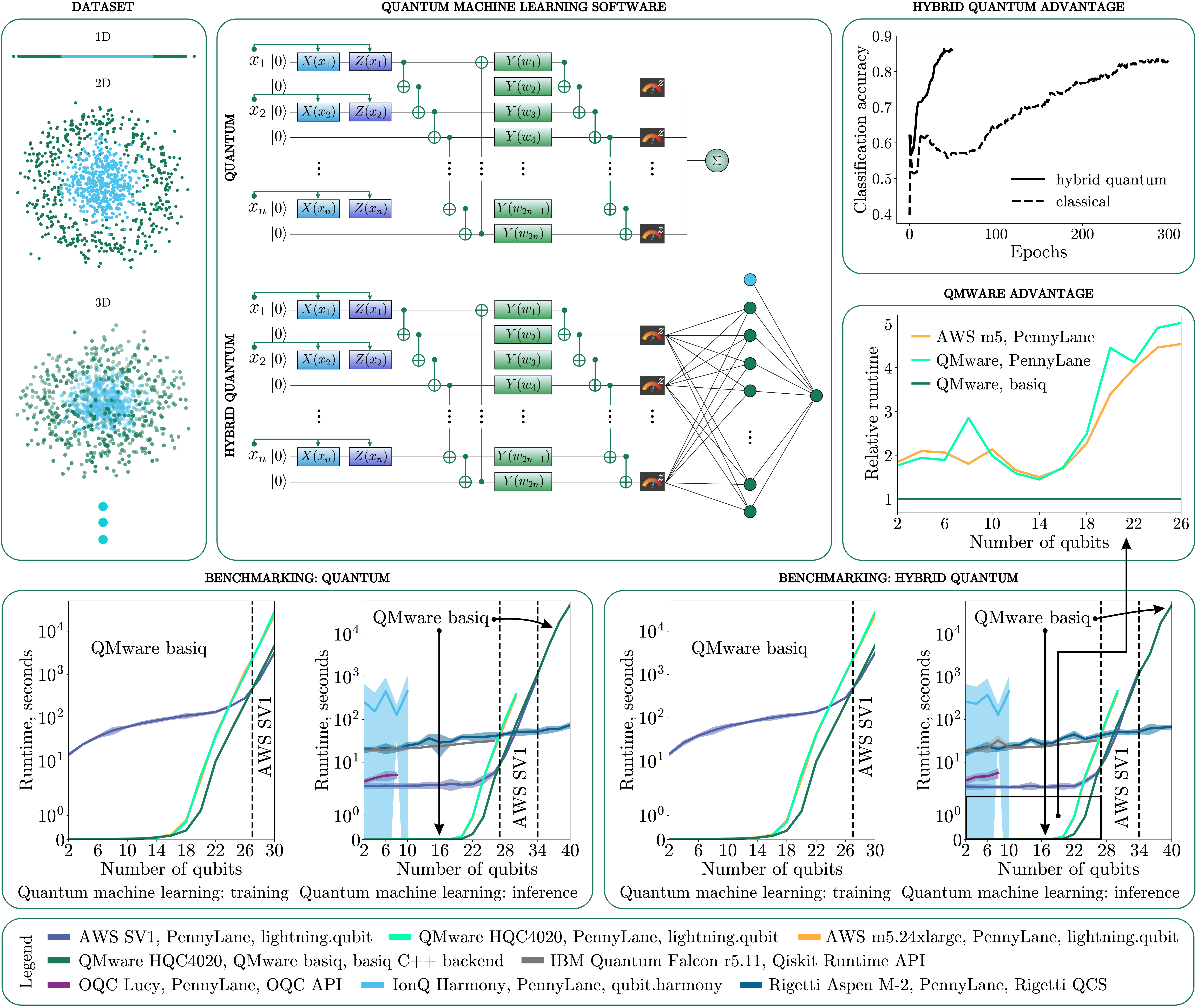}
    \caption{Overview of benchmarking methodology and results. Boxes clockwise from top left: examples of the $n_d$-dimensional dataset used for benchmarking for $n_d = 1, 2, 3$, each containing $m = 500$ data points; the architecture of the pure quantum (top) and hybrid (bottom) neural networks used for the benchmarking calculations; the classification accuracy of the HQNN and an equivalent classical neural network, illustrating the improved learning rate of a hybrid approach; the relative runtime of different software and hardware stacks for a range of circuit sizes, showing the performance advantage of the basiq SDK using QMware's cloud computing service; training (left) and inference (right) runtime for the HQNN on various simulator software and hardware stacks; training (left) and inference (right) runtime for the pure QNN on various simulator software and hardware stacks (see legend at bottom). Dashed lines in the runtime plots illustrate the most performant stack for a given range of qubits. The inference plots in both cases also illustrate the runtime using real QPUs.}
    \label{fig:benchmarking-overview}
\end{figure*}

\begin{table}[h]
    \centering
    \begin{tabular}{|c|c|c|}
        \hline
        Hardware & Qubits & Native Gates\\
        \hhline{|=|=|=|}
        Rigetti Aspen M-2 & 80 & $\mathrm{RX}$ $\mathrm{RZ}$ $\mathrm{CZ}$, $\mathrm{CP}$\\
        IBMQ Falcon r5.11 & 27 & $\mathrm{I}$, $\mathrm{CX}$, $\mathrm{IFELSE}$, $\mathrm{RZ}$, $\mathrm{SX}$, $\mathrm{X}$ \\
        IonQ Harmony & 11 & $\mathrm{GPI}$, $\mathrm{GPI2}$, $\mathrm{MS}$\\
        OQC Lucy & 8 & $\mathrm{I}$, $\mathrm{ECR}$, $\mathrm{V}$, $\mathrm{X}$, $\mathrm{RZ}$ \\
        \hline
    \end{tabular}
    \caption{Quantum processing units used in the hardware benchmarking tests, their qubit counts, and native gates.}
    \label{tab:benchmark-qpus}
\end{table}

Initially the performance of the QMware HQC4020 simulator is compared to the performance of the AWS Braket ml.m5.24xlarge simulator. In both instances the PennyLane \texttt{lightning.qubit} backend is used in order to evaluate the performance of the underlying hardware systems -- referred to as QPL and APL, respectively. The performance of the QMware basiq library is then benchmarked using both a native basiq implementation of the QNN and HQNN. This runtime benchmark is executed on the QMware HQC4020 simulator using 384 vCPUs across all circuit sizes. Results are compared to the high-performance AWS Braket SV1 simulator (ASV), as well as the previous QPL and APL benchmark. Finally, the runtime performance of these simulator stacks is also compared to runtimes achieved for (H)QNN inference (forward pass) using real QPUs. The QPUs included are IonQ's Harmony, OQC's Lucy, Rigetti's Aspen M-2, and IBM Quantum's Falcon r5.11. 

The ml.m5.24xlarge AWS notebook instance provides 96 vCPUs and 384 GiB of random access memory (RAM). By comparison the QMware HQC4020 simulator has 12 TB of RAM and 384 vCPUs in total, a maximum of 48 vCPUs of which are utilized throughout the benchmarking tests executed on the QMware service. The exception to this is when benchmarking the PennyLane Lightning qubit, which implements no parallelization for the parameter-shift method. Hence the results for the QMware PennyLane \texttt{lightning.qubit} (QPL) and AWS PennyLane \texttt{lightning.qubit} (APL) benchmark represents single-core calculations on both hardware services.

The metric used in benchmarks is the training time per epoch per training sample, which is measured using the Python \texttt{time} library. The expectation value of each circuit measurement, corresponding to the output prediction of the (H)QNN, is obtained using 1000 circuit shots. Measuring the execution time for multiple circuit shots has the effect of reducing the proportion of  circuit initialization time in the quoted runtime for each stack. In practice, quantum circuits usually require hundreds or thousands of circuit shots to obtain an accurate expectation value. Benchmarks are performed for a range of circuit sizes (number of qubits). This is achieved by varying the dimension of the dataset, $n_d \in [1, 15]$, which increases linearly with the number of qubits, $n_q = 2 n_\text{d}$ (see Sec.~\ref{sec:runtime-benchmarking-dataset} for details).

Although the IBM Quantum Cloud development platform allows users to retrieve quantum processor runtimes, the AWS Braket computing platform precludes the possibility of measuring low-level process times. As a result all times quoted, for both AWS and QMware benchmark, include the interaction with the runtime environment and the backend that compiles, initializes, and invokes the quantum circuit. This follows the standard practice in quantum benchmarks and reflects a real-world use case encompassing the full software and hardware stack~\citep{wac_qua_2021}. For the benchmark involving the QPU devices available via AWS Braket, quoted times also include any potential queue time incurred on the vendor backend. The proportion of the total runtime this constitutes varies according to the number and complexity of tasks submitted by other users of the same QPU (as detailed in the Amazon Braket Developer Guide) and cannot be accurately estimated through the AWS user interface.

\subsection{Benchmark Dataset}
\label{sec:runtime-benchmarking-dataset}

The dataset for the benchmarks presented here is an $n$-dimensional abstracted version of the well-documented two-dimensional \texttt{sklearn.datasets.make\_circles} for binary classification tasks~\citep{ped_sci_2011}. It consists of points sampled from two concentric circular shells, distributed randomly about two nominal mean radii, $r_\text{inner} < r_\text{outer}$. The distribution of the data points about the mean radius is described by a normal distribution, and both shells use the same standard deviation. The method for creating the dataset is adapted from that proposed by \citep{mar_cho_1972} for sampling points on a sphere. First, an $n_d$-dimensional vector is created with components $v_i$ sampled randomly from a standard normal distribution with mean $\mu = 0$ and standard deviation $\sigma = 1$:
\begin{equation}
    \vec{v} = \begin{bmatrix}
        v_1 \\
        v_2 \\
        \vdots \\
        v_{n_d}
    \end{bmatrix}, \quad  v_i \in \mathcal{N}(0, 1).
\end{equation}
The vector is then normalized to length $r$ to obtain a point sampled randomly with uniform probability from the surface of an $n$-sphere. A vector of random noise, $\vec{\rho}$, sampled independently from the distribution $\mathcal{N}(0, \sigma^2)$, is applied to each component of the vector:
\begin{equation}
    \vec{x} = r\frac{\vec{v}}{\norm{\vec{v}}} + \vec{\rho}, \quad \rho_i \in \mathcal{N}(0, \sigma^2).
\end{equation}
Data points $\vec{x}$ are sampled from two such distributions to create an outer shell with classification label $y_i = 0$ and inner shell with classification label $y_i = 1$. In the present work $r_\text{outer} = 1.0$ is used to create points in the outer shell and $r_\text{inner} = 0.2$ to create points in the inner shell, with $\sigma = 0.3$ for both shells. The dimension of the dataset determines the number of features used as input to the neural network. By linearly increasing the dimension of the dataset, circuits with a varying number of qubits ($n_q = 2n_d$) can be benchmarked without changing the underlying rubric of the classification problem itself. Examples of the one-, two- and, three-dimensional datasets are shown in Fig.~\ref{fig:benchmarking-overview}.

\subsection{Learning Models}
\label{sec:learning-models} 
The following sections describe the architecture of the hybrid quantum-classical (HQNN) and pure QNNs used in the benchmarking tests. In all cases, the networks are trained using a binary cross-entropy loss function and the Adam optimizer with a learning rate of $\alpha$ = 0.3~\citep{kin_ada_2017}.

\subsubsection{Quantum Neural Network}
\label{sec:quantum-neural-network} 

The QNN used in this benchmark consists of a multi-qubit variational quantum circuit, and is based on the model proposed by \citep{per_pra_2022} The model employs a sequential \RX\RZ two-axis rotation encoding scheme to embed each of the data features in a single qubit state on the odd-numbered qubits. The feature encoding is followed by an entangling layer of sequential `nearest-neighbor' CNOT gates across all qubits. A layer of trainable single-axis \RY rotation are then applied to each qubit, followed by a final layer of sequential `cascading' CNOT gates across all qubits. Finally, the expectation value of the Pauli-Z operator ($\sigma_z$) is measured for each of the even-numbered qubits. The mean expectation value across the $n_d$ measurement qubits is interpreted as a probability of the input belonging to the class $y_i = 1$.

To obtain the gradients of the loss function with respect to each of the parameters, the standard analytical \textit{parameter-shift} algorithm is used, in which the gradient of an expectation value, with respect to the $i$-th parameter, is obtained via measurement of two additional expectation values. This method requires a total of $2n_w + 1$ circuit evaluations to obtain the gradient of the loss with respect to $n_w$ circuit parameters and thus scales linearly with the number of trainable parameters. Unlike the more efficient \textit{adjoint} and \textit{backpropagation} methods, which are usually available for quantum simulators, parameter-shift is the only currently available algorithm that can be implemented natively on a QPU. It therefore provides an upper bound on the cost of training a QNN using a QPU, and on the runtime of the benchmarked simulators~\citep{rum_lea_1986, jon_eff_2020}.

\subsubsection{Hybrid Quantum Neural Network}
\label{sec:hybrid-neural-network}

The HQNN is a bipartite model consisting of a QNN and a multi-layer perceptron (MLP) classical neural network, with the expectation value of each measurement qubit in the QNN used as an input feature for the first layer of the MLP. The quantum component of the HQNN follows the same architecture as described in Sec.~\ref{sec:quantum-neural-network}, and uses the same parameter-shift method to obtain the gradients of the expectation values. The MLP is built using the PyTorch library and contains three linear layers with sizes $[n_d, 40, 1]$, with  ReLU and Sigmoid activation functions applied to the input and hidden layer, respectively~\citep{pas_pyt_2019}. The gradients in the classical component of the HQNN are computed using the standard back-propagation algorithm, via the native implementation in the PyTorch library. For the inference benchmark that utilizes QPUs or the AWS SV1 device, the classical part of the circuit is executed using the AWS ml.m5.24xlarge compute instance in all cases except for IBM Quantum Falcon r5.11 whose classical part was executed on QMware HQC4020.

\subsection{Results}
\label{sec:results}

This section presents the results of the benchmark methodology outlined in Sec.~\ref{sec:methodology}. The results of the benchmark are illustrated in Fig~\ref{fig:benchmarking-overview}. A table of these results is also given in App.~\ref{sec:benchmarking-results}, Tab.~\ref{tab:data-training} and Tab.~\ref{tab:data-inference}. The measured runtime values (runtime per epoch, per training sample) are averaged across 100 repeats in order to obtain a mean and standard deviation for circuits up to 24 qubits in size for the training benchmark, and 26 qubits for the inference benchmark. A similar approach to averaging is impractical for larger circuit sizes, owing to the significantly longer total runtime. Thus, only a single value with no standard deviation is quoted for circuits larger than 24 qubits. In all cases Chauvenet's criterion is applied in order to filter anomalous runtime measurements that arise due to extraneous hardware processes ~\citep{chauvenet_1863_a}. 

\begin{table}[h]
    \centering
    \begin{tabular}{|c|c|c|c|}
        \hline
        Label & Hardware & SDK & Backend  \\
        \hhline{|=|=|=|=|}
        QBN & QMware HQC4020 & basiq & basiq C++ \\
        QPL & QMware HQC4020 & PennyLane & lightning.qubit \\
        APL & AWS ml.m5.24xlarge & PennyLane & lightning.qubit \\
        ASV & AWS SV1 & PennyLane & SV1 \\
        \hline
    \end{tabular}
    \caption{The abbreviations associated with each hardware, SDK, and backend. }
    \label{tab:benchmark-labels}
\end{table}

\subsubsection{Training}
\label{sec:training}

In general, the QMware HQC4020 and the AWS ml.m5.24xlarge hardware achieve similar performance using the Pennylane Lightning backend (QPL and APL, respectively - see Tab.~\ref{tab:benchmark-labels} for a list of abbreviations). When benchmarking with the QNN, QPL performs similarly to APL for circuits with less than 27 qubits in size, with a relative runtime of $t_\text{QPL}/t_\text{APL}$ = \num{0.90 \pm 0.17}. When benchmarking with the HQNN the average relative runtime is $t_\text{QPL}/t_\text{APL}$ = \num{0.95 \pm 0.16}.

Comparing the QMware HQC4020 hardware using a native basiq implementation (QBN) to the Pennylane Lightning implementation (QPL) shows a clear advantage for the native basiq implementation across all circuit sizes for both QNNs and HQNNs. The QBN implementation achieves an average speedup of $t_\text{QBN}/t_\text{QPL}$ = \num{0.56 \pm 0.32} over QPL in the QNN benchmark, and $t_\text{QBN}/t_\text{QPL}$ = \num{0.54 \pm 0.30} for HQNNs. It is also notable that the native QMware implementation is most performant for models using fewer than more than 20 qubits, where an average speedup of $t_\text{QBN}/t_\text{QPL}$ = \num{0.22 \pm 0.04} is obtained for the QNNs and $t_\text{QBN}/t_\text{QPL}$ = \num{0.21 \pm 0.02} for the HQNNs.

The AWS Braket SV1 device is a high-performance managed device designed for simulating large quantum circuits up to a maximum of 34 qubits. Correspondingly, it outperforms QBN for circuits with 28 or 30 qubits with an average relative runtime across QNNs and HQNNs of $t_\text{QBN}/t_\text{ASV}$ = \num{1.35 \pm 0.11}. Conversely, it has very poor performance for small and medium circuits, with a relative runtime approximately two to four orders of magnitude slower than the other available simulator stacks for circuits with fewer than 20 qubits. Across all circuit sizes smaller than 28 qubits, the QBN implementation outperforms ASV for both QNNs and HQNNs. The average relative runtime is $t_\text{QBN}/t_\text{ASV}$ = \num{0.00041 \pm 0.00025} for circuits with 14 qubits or less, $t_\text{QBN}/t_\text{ASV}$ = \num{0.0049 \pm 0.0035} for circuits with 16 to 20 qubits, and $t_\text{QBN}/t_\text{ASV}$ = \num{0.36 \pm 0.29} for circuits with 22 to 26 qubits. 

\subsubsection{Inference}
\label{sec:inference}

When training a QNN the number of circuit evaluations increases linearly with the number of trainable parameters. In contrast, the inference (forward pass) requires only a single evaluation of the output expectation values. As a result, the total runtime is reduced significantly. The results of the inference runtime benchmark broadly follow the same trends as for the training runtime. One exception to this are the QPL and APL relative runtimes, which are identical within one standard deviation for circuits less than 16 qubits in size. The average relative runtime is $t_\text{QPL}/t_\text{APL}$ = \num{1.04 \pm 0.10} for QNNs and $t_\text{QPL}/t_\text{APL}$ = \num{1.03 \pm 0.22} for HQNNs. For larger circuits, APL performs marginally better than QPL with $t_\text{QPL}/t_\text{APL}$ = \num{1.11 \pm 0.10} and $t_\text{QPL}/t_\text{APL}$ = \num{1.12 \pm 0.09} for QNNs and HQNNs, respectively.

Crucially, the inference tests include a benchmark of physical QPUs, as listed in Tab.~\ref{tab:benchmark-qpus}. For low numbers of qubits the QPU runtimes are orders of magnitude longer than their simulator counterparts. This is likely due to the additional overhead incurred in compiling and initializing a QPU results, as well as the queue time incurred on the vendor side, where multiple AWS users may be accessing the QPU device simultaneously. On the other hand, the runtime for QPUs increases linearly with the number of qubits. For large numbers of qubits, the exponentially increasing simulator runtime exceeds the fixed time cost associated with QPUs. This results in a `threshold' circuit size above which it becomes exponentially faster to execute a QNN using a QPU device.

The limited number of qubits available for many QPUs means that, in most cases, this threshold is not attainable. Of the three QPUs in the present study, only Rigetti's Aspen M-2 has a sufficient number of qubits to be time-competitive with the simulator stacks tested. The threshold occurs at 30 qubits for QBN and ASV where the relative runtime is $t_\text{ASV}/t_\text{M-2}$ = 1.03, $t_\text{QBN}/t_\text{M-2}$ = 1.53 for the QNN and $t_\text{ASV}/t_\text{M-2}$ = 1.02, $t_\text{QBN}/t_\text{M-2}$ = 1.51 for the HQNN. For smaller circuits sizes OQC's Lucy produces runtimes that are faster by a factor of $t_\text{Lucy}/t_\text{M-2}$ = \num{0.22 \pm 0.03} compared to Rigetti's Aspen M-2. In contrast, the runtime for IonQ's Harmony QPU is on average a factor $t_\text{Harmony}/t_\text{M-2}$ = \num{13.5 \pm 5.2} slower than Aspen M-2. The IBMQ Falcon r5.11 QPU performs similarly to the Rigetti Aspen M-2, with an average relative runtime of $t_\text{ASV}/t_\text{M-2}$ = \num{0.87 \pm 0.18}. 

Notably, vendor management of the QPUs results in a runtime that varies significantly from job to job. The percentage variance in runtime measured for inference on a QPU is generally higher than for simulator execution. Specifically, the average percentage standard deviation across the 10 repeat measurements is on the order of 125\%, 16\%, 21\%, and 7\% for IonQ, Aspen M-2, Lucy, and Falcon r5.11, respectively.

\section{Accuracy and Cost Analysis}
\label{sec:accuracy-and-cost-benchmarking}

\begin{figure*}[th!]
    \centering
    \includegraphics[width=0.9\textwidth]{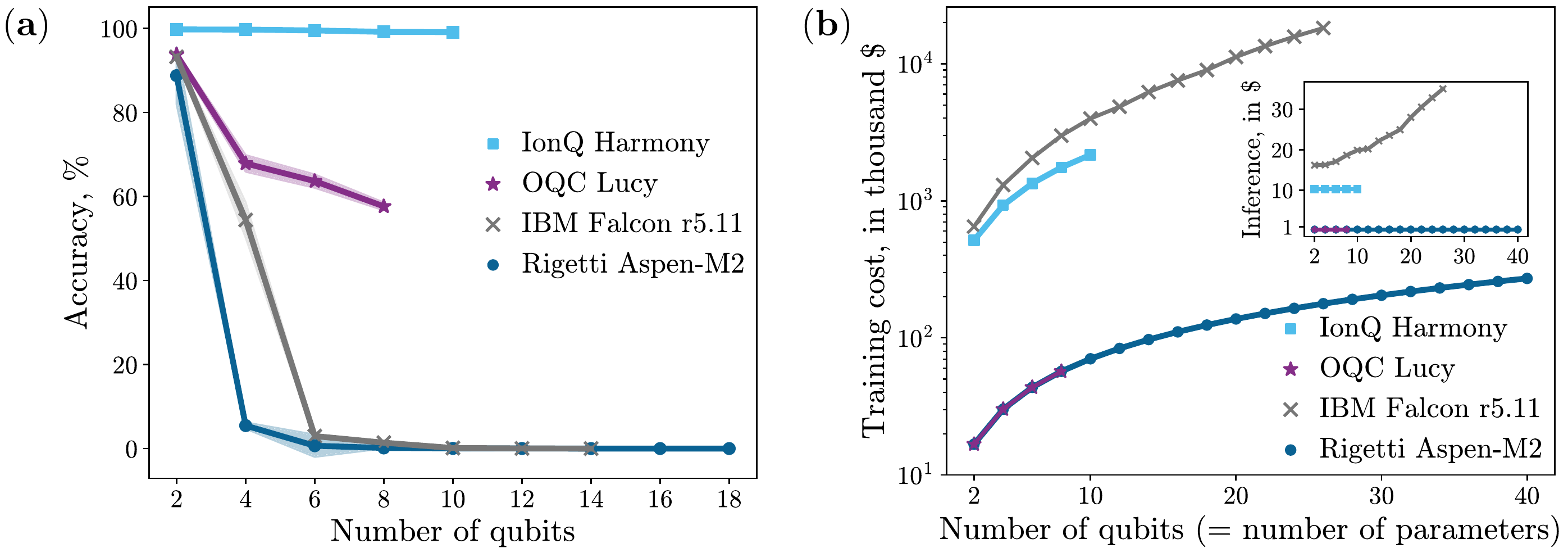}
    \caption{Comparison of circuit fidelity (left), training cost (right) and inference cost (inset) for various publicly available QPUs. The training price estimates are calculated on the assumption that the model is trained over 100 epochs, with 100 training samples and 1000 circuit shots per expectation value. For IBM Falcon r5.11 the prices are obtained in CHF and the conversion is calculated using a rate of 1.06 USD = 1.00 CHF. Note that the IBM Cloud Estimator service offers a method for circuit training that could reduce the IBM prices by up to 2 orders of magnitude, please refer to Sec.~\ref{sec:cost_eval} for a detailed explanation. }
    \label{fig:accuracies-price}
\end{figure*}

\subsection{Cost Evaluation}\label{sec:cost_eval}

As described in Sec.~\ref{sec:introduction} a primary consideration in developing and testing QNNs and HQNNs is the financial cost of training a network. In general, the pricing structure of publicly available QPUs is such that the training cost is proportional to the number of distinct quantum circuits that must be evaluated during the training process. When using the parameter-shift method for gradient calculations, the number of distinct quantum circuits is proportional to the number of trainable parameters. Hence, for the QNN and HQNN considered in Sec.~\ref{sec:quantum-neural-network} and Sec.~\ref{sec:hybrid-neural-network}, the training costs scales linearly with the number of qubits. For an increasing number of epochs, and for an increasing number of training samples, this rapidly results in millions or billions of distinct quantum circuits that must be initialized and evaluated on the QPU. 

In the present work, QPUs are accessed through AWS Braket and IBM Cloud Qiskit Runtime. The AWS pricing scheme includes both a per-task cost, incurred for the execution of a given quantum circuit, and a per-shot cost which is applied to the number of shots specified for that quantum circuit. By comparison, Qiskit Runtime implements a pricing scheme on the basis of runtime with a fixed price per second for executing a quantum circuit.  Fig.~\ref{fig:accuracies-price}(b) illustrates an example of how the estimated cost scales with circuit size for the four QPUs presented in this work. The consequence of this QPU pricing structure is that for circuits larger than a few bits, training a QNN on a QPU becomes prohibitively expensive. Prices range from approximately 1000 USD for a two-qubit QNN using Rigetti's Aspen M-2 or OQC's Lucy, to more than \num{10e6} USD for a 26 qubit QNN using IBM's Falcon r5.11.  It is worth clarifying that the training in this benchmark treats every quantum evaluation as a new circuit on which many shots can be executed. The IBM Qiskit Runtime offers an alternative method: where the quantum computer scientist can set up the quantum circuit once for a large initial cost and then all the circuit executions in that epoch would use the same setup but with varying parameters depending on the dataset and the parameter-shift rule. Assuming a circuit set-up that takes $t_1 = 5s$ and a per shot runtime of $t_2 = 250 \mu s$, the cost of training the same circuit could be dramatically reduced from tens of millions of USD to $212,800$ USD. The multi-parameter, multi-circuit functionality is specific to the IBM Cloud and thus the authors decided to set up every circuit in all cases in the interest of fairness.

In contrast to the high cost of training, the forward pass of a QNN does not entail a gradient calculation and thus requires only a single task with around 100-1000 circuit shots. The cost of using a QPU for the inference stage of a trained QNN is considerably less. For the QPUs accessed through AWS Braket, the cost is fixed for a given number of shots, irrespective of circuit size. For QPUs accessed through Qiskit Runtime, the cost increases linearly with circuit size (QPU runtime increases linearly with circuit size). A typical forward-pass of a QNN can be executed at a cost of approximately 1 USD for QPUs accessed through AWS Braket, or between 10 -- 40 USD for QPUs accessed through Qiskit Runtime. 

\subsection{Accuracy Evaluation}
\label{sec:accuracy}

Understanding the role noise plays in executing quantum circuits is critical to leveraging quantum computers. Vendors often provide esoteric measures of fidelity, gate accuracy and characteristic timescales. Although these are valuable metrics for assessing the relative performance of different QPUs, evaluating the overall effect of these different sources of error on a typical quantum circuit can be challenging. A holistic measure of noise in a quantum circuit can be achieved with a straightforward empirical procedure. 

First, the parameters of all trainable gates are fixed at a value of $\pi/4$. The input feature values are each set to $\pi^2/4$, and the QNN is augmented by applying the adjoint of the entire circuit prior to measurement. In a noiseless circuit this results in a final state vector with zero amplitude for all basis states except the computational ground state, i.e $\ket{00...0}$. In a physical QPU, various noise sources degrade the fidelity of the circuit, resulting in a non-zero probability of measuring one of the other $2^{n_q}-1$ computational basis states. An accuracy measure is then obtained by counting the proportion of states measured in the computational ground state over 1000 circuit shots. This is commonly referred to as the fidelity of a quantum state, $F = \left|\braket{00...0}{\psi}\right|^2$. In this case the fidelity of the final quantum state is measured relative to the computational ground state. This fidelity measurement is repeated by executing 10 such jobs on each QPU to obtain a mean accuracy. The state vectors needed to obtain fidelities for individual shots are not available through the PennyLane SDK with AWS. Consequently, the AWS braket software library is used instead to construct the same quantum circuits described in Sec.~\ref{sec:learning-models} for physical QPU accuracy tests presented here, with the exception of the IBM Falcon r5.11 which was tested on QMware and accessed through IBM Cloud's Qiskit Runtime API. 

The results, shown in Fig.~\ref{fig:accuracies-price}(a) and Tab.~\ref{tab:data-qpu-accuracies}, vary significantly between QPUs. The IonQ Harmony device attains greater than 99\% fidelity across circuits sizes from two to ten qubits. The fidelity of the OQC Lucy device is high for small circuits with only two qubits, but degrades significantly for larger circuits, with $67 \pm 2$ \%, $64 \pm 2$ \% and $57.6 \pm 0.9$ \% fidelity for four, six, and eight qubits, respectively. The performance of the Rigetti Aspen M-2 and IBMQ Falcon r5.11 QPUs is similarly high for the two qubit circuit, with fidelities of $89 \pm 6$ \% and $93.2 \pm 0.4$ \%, respectively. However the fidelity of these devices approaches zero for circuits larger than approximately eight qubits.  

Note that the IonQ Harmony device has full connectivity between its 11 qubits, potentially resulting in a shallower transpilation depth that could contribute to its higher overall accuracy. Importantly, there are additional considerations that might contribute to this result, such as internal optimizations or error correction algorithms.  Since this study focuses on performing a practical benchmark, these results were obtained from the end-user's perspective with minimal modification or tuning. Thus, this work does not rule out potential gains for other devices if the appropriate tuning is performed. 

\section{Discussion and Summary}
\label{sec:discussion}

\begin{figure*}
    \centering
    \includegraphics[width=0.9\textwidth]{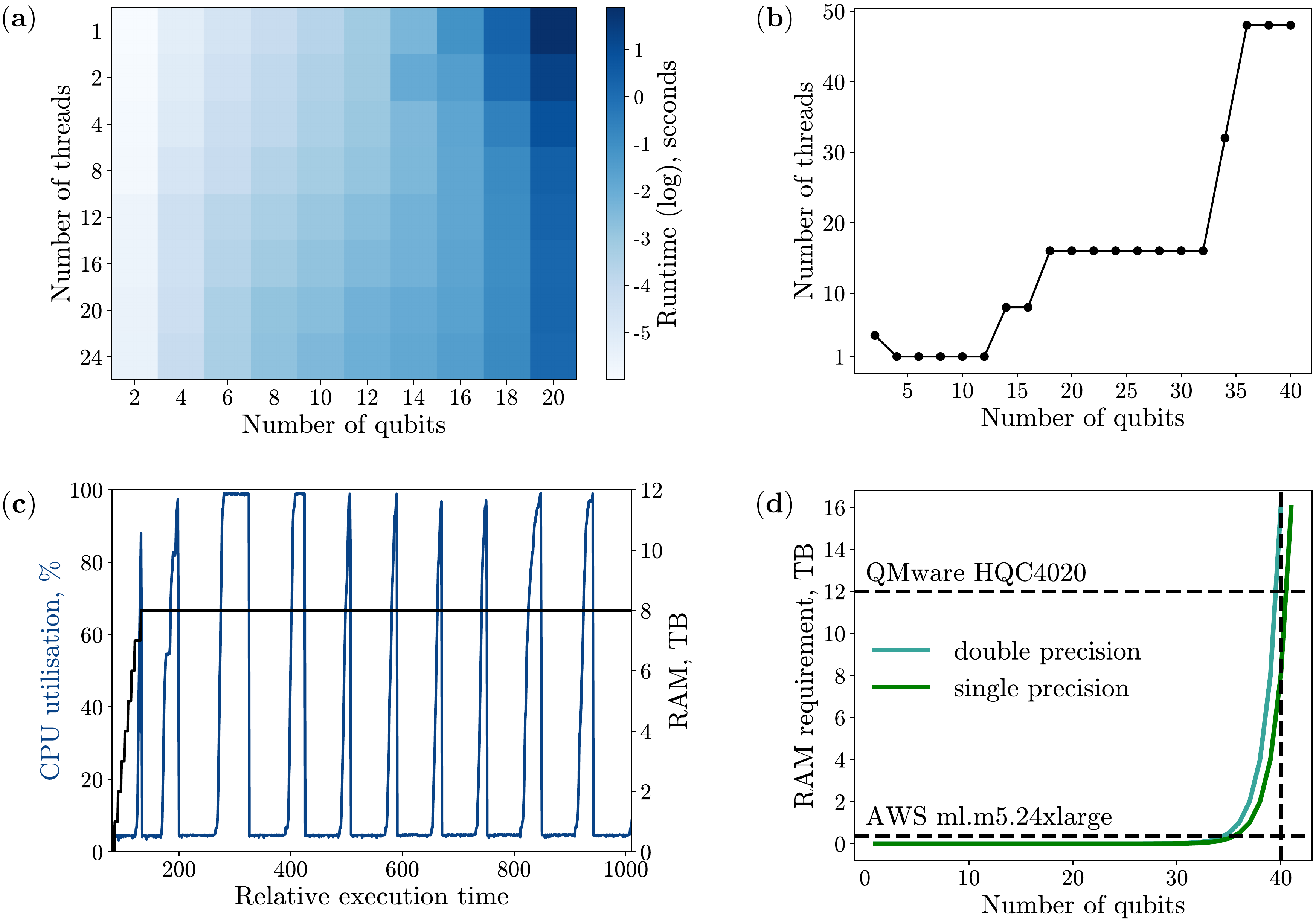}
    \caption{Hardware statistics for the QMware HQC4020 simulator. (a) The optimal number of threads for simulating an $n$-qubit quantum circuit. Faster runtimes are indicated with a paler shade, demonstrating how the optimal number of threads varies for increasing circuit size. (b) a plot of the optimum thread count against number of qubits (c) a plot of total CPU and RAM utilization over time during the execution of a 40 qubit QNN. (d) memory required to store the state vector for an increasing number of qubits.}
    \label{fig:hardware-utilization}
\end{figure*}

\subsection{Large Quantum Circuits}
\label{sec:large-quantum-circuits}

The amount of random access memory (RAM) utilized in simulating any noiseless quantum circuit is a function of the dimension of the vector space, and hence grows exponentially with the number of qubits. An $n$-qubit state is specified by $2^n$ complex amplitudes, and thus requires approximately $16 \times 2^n$ bytes of memory. This equates to approximately 16 GB of RAM for a 30 qubit circuit.

QNNs using large numbers of qubits are highly susceptible to the barren plateau phenomenon, where the gradients of a QNN with randomly initialized parameters vanish exponentially as the number of qubits increases~\citep{mcc_bar_2018}. This could present a serious obstacle to training a QML model with a large number of qubits unless a variety of mitigation methods are employed~\citep{zha_ana_2021, vol_lar_2021, gra_ini_2019,kordzanganeh2022exponentially}. Solving the barren plateau problem is crucial to achieving highly expressive models that could outperform classical machine learning methods on complex datasets. Access to development services that are able to simulate QNNs with a large number of qubits is essential to solving the barren plateau problem.

To explore the performance of QMware's HQC4020 with an increasing number of qubits, additional benchmarks are performed for QNNs and HQNNs with up to 40 qubits. A 40 qubit simulation is achievable by reducing the floating point precision to single precision (32 bit) representation, for all other circuit sizes a double precision (64 bit) is retained. Fig.~\ref{fig:hardware-utilization}(d) illustrates the exponential increase in memory usage for a range of circuit sizes up to 40 qubits, for both single and double precision representations. 

Tab.~\ref{tab:data-large-circuit-runtime} gives the inference runtimes for circuits with up to 40 qubits, up to a maximum of 34 qubits in the case of ASV. The runtimes for the QBN and ASV simulators increase exponentially, with QBN reaching a runtime greater than 13 hours for the 40 qubit QNN. Owing to the long simulator runtimes for large circuits, multiple repeats are not possible, and thus it is hard to draw meaningful conclusions from the single QBN and ASV trials presented for circuits larger than 27 qubits. Nonetheless, in the case of QNNs, ASV achieves a relative runtime of $t_\text{ASV}/t_\text{QBN}$ = \num{1.34 \pm 0.09} for circuits with 28 to 34 qubits. In the case of HQNNs there is no clear advantage, with $t_\text{ASV}/t_\text{QBN}$ = \num{0.95 \pm 0.22}.

\subsection{Multi-threading}
\label{sec:multi-threading}

The runtime performance for circuits with many qubits can be improved using various parallelization techniques. A common method for achieving a substantial increase in runtime performance for linear algebra operations is to compute matrix-vector and matrix-matrix products with the aid of multi-threading. The QMware basiq library provides native support for a multi-threaded approach to low-level C++ linear algebra operations. However, determining the optimal number of threads to execute a circuit with a given number of qubits is not straightforward. Fig.~\ref{fig:hardware-utilization}(a) illustrates a cross-sectional study of runtime for circuits with up to 20 qubits with multi-threading across as many as 24 threads. In Fig.~\ref{fig:hardware-utilization}(b) the optimum number of threads for a given qubit count is shown. These results can be applied generally to achieve best-in-class performance with the QMware HQC2040 using the basiq software library.

PennyLane provides some parallelization for gradient calculations using the adjoint operator method \citep{jon_eff_2020}, which is applicable to parameterized quantum algorithms. However, the authors are not aware of any low-level parallelization in the PennyLane library for general linear algebra calculations encountered in generic quantum circuits.

\subsection{Conclusion}
\label{sec:summary}

This work presents a comprehensive study of various quantum computing platforms, using both simulated and physical quantum processing units. The results presented in Sec.~\ref{sec:results} demonstrate a clear runtime advantage for the QMware basiq library executed on the QMware cloud computing service. Relative to the next fastest classical simulator, QMware achieves a runtime reduction of up to 78\% across all algorithms with fewer than 27 qubits. In particular, the QMware basiq library achieves a runtime reduction of \num{0.56 \pm 0.32} relative to the PennyLane library for QNNs and \num{0.54 \pm 0.30} for HQNNs. 

The QMware HQC4020 hardware benchmarked against AWS ml.m5.24xlarge achieves a comparable relative runtime of \num{0.90 \pm 0.17} for QNNs and \num{0.95 \pm 0.16} for HQNNs. Thus, the advantage offered by the QMware cloud computing service is primarily due an harmonious interplay between software and hardware. This performance advantage can be attributed to the superior multi-threading support present in the basiq library. AWS also provides the SV1 simulator, which performs marginally better than QMware basiq for large circuits with more than 27 qubits in the case of QNNs (up to the SV1 maximum of 34 qubits). There is no clear advantage for either QMware or SV1 in the case of HQNNs. Additionally, QMware is the only tested simulator that has the capability of simulating circuits with 34-40 qubits.

The price and scarcity of quantum hardware means it is more time and cost efficient to develop algorithms and train QNNs with quantum simulators such as Amazon Web Services Braket or QMware's cloud computing service. This is particularly true for variational quantum algorithms and QNNs which represent a promising utilization of quantum technologies in artificial intelligence on NISQ computers. In contrast to the exponential runtime scaling encountered in quantum simulators, the runtime of QPUs scales linearly with the circuit size. Publicly available QPUs are already able to achieve runtime improvements over simulator hardware for large numbers of qubits. For example, Rigetti's Aspen M-2 is able to execute 40 qubit circuits in approximately one minute, which is less than 0.02\% of the runtime measured for QMware's simulator.

As quantum hardware improves and the number of qubits available grows, it will become possible to gain a substantial runtime advantage over simulator hardware when executing large quantum circuits. However, the fidelity tests presented in Sec.~\ref{sec:accuracy} indicate that accurate inference with such a large quantum circuit is not yet possible. Moreover, the cost of accessing these quantum devices makes training a QNN on currently available QPUs prohibitively expensive. Owing to the exponential computational space, QNNs with relatively few qubits are able to tackle challenging data science and industry problems. Thus, the key to success in the field of quantum computing is to improve the cost and the accuracy of QPUs, and integrating them well within classical infrastructure. The hybrid interplay between quantum and classical machines is the key to seamlessly harness the best performance of QPUs and simulators depending on the use case. 

\subsection{Author contributions}
M.K., M.B., Maxim P., W.F., A.K., W.S., and A.S. wrote the benchmarking Python code. B.K. reviewed the existing benchmarking literature, wrote the Appendix, and automated the benchmarking Python code. M.K., Maxim P., and A.S. performed benchmarking of simulated quantum processing units. M.K., A.K., and W.S. performed benchmarking of physical quantum processing units. A.M., M.K., W.S., A.K., and A.S. analyzed the results. M.B. and W.F. set up the QMware simulator, and prepared the basiq SDK.
M.B. monitored the resource utilization on the QMware simulator, and analyzed the multithreading efficiency. W.S. and M.K., and A.M. wrote the initial version of the main text. A.S. and A.M. prepared the figures. All authors have contributed to improving the manuscript, read, and agreed to the final version of the manuscript. Markus P. and A.M. performed project administration and supervision.

\bibliography{main.bib}
\bibliographystyle{unsrt}


\clearpage

\appendix

\section{Current Benchmarking Landscape} \label{App:BM}

    As described in Ref. \cite{lan_2022_rac}, quantum benchmarking faces many more challenges than its classical counterparts. After all, there is no singular ``quantum technology.” Quantum devices are made from different materials and used for vastly different purposes. This makes it difficult to predict how the field will progress, as each type of technology develops breakthroughs at different rates. Additionally, quantum devices are ultimately compared not only to other quantum computers but also to the performance of classical devices, itself an ever-moving target.

\begin{figure*}[t]
    \centering
    \includegraphics[width=0.75\textwidth]{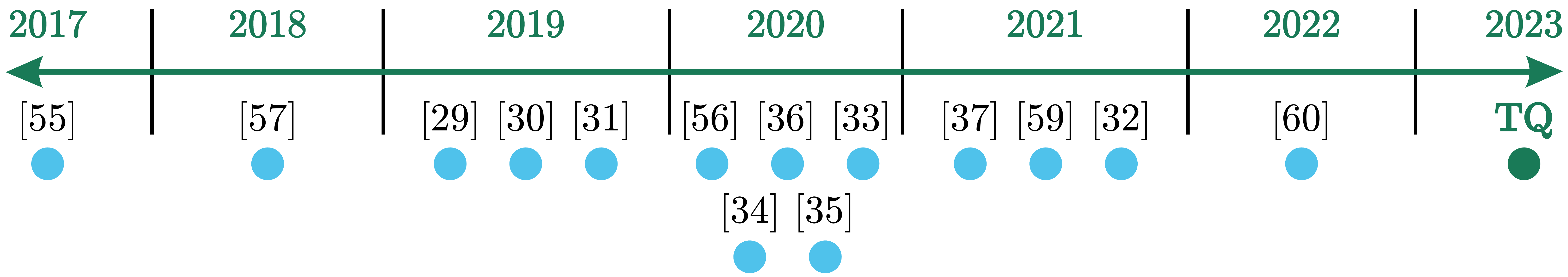}
\caption{
    Timeline of key benchmarking papers considered in Appendix: Current Benchmarking Landscape. The bracketed numbers represent the citations of the referenced papers. 
}
    \label{fig:bm_timeline}
\end{figure*}

    Given the multitude of challenges, it is important to question how any benchmark result compares to others. In this appendix, a non-exhaustive list of key takeaways from other benchmarking papers is presented to better address where this paper sits in the wider landscape. 
    The year of each paper’s publication can be viewed in Figure \ref{fig:bm_timeline}. 
    For more information on the current state of quantum benchmarking, Ref. \cite{wan_2022_sok} provides an excellent summary of many benchmarking papers. In that paper, Wang et al. separate quantum benchmarks into three classes: physical, aggregative, and application-level.

    Physical benchmarks depend on hardware specificity. These tasks give a measure of the engineering capabilities and limitations of quantum devices, such as the impacts of decoherence and noise, which have been established as major speedbumps in the development of fully-functional quantum devices \cite{mic_2017_ben,sal_2019_ben}. Examples of physical benchmarking papers include Ref. \cite{wri_2019_ben}, a measurement of qubit coherences and gate-based fidelities of IonQ's trapped Ytterbium qubits, and Ref. \cite{koc_2020_dem}, which measures T1 and T2 coherence times as well as the CCNOT gate and the adjoint Quantum Fourier Transform for a 20-qubit IBM device.   

    Aggregated benchmarks measure how well multiple qubits and gates work together. Perhaps the most widely used of these hardware-agnostic metrics is Quantum Volume. Introduced by IBM in Ref. \cite{cro_2019_val}, Quantum Volume measures the maximum circuit width (number of qubits) and depth (layers of gates) that a device can run such that the two numbers are equal and the fidelity is maintained above a certain threshold. Given a square circuit of width and depth size $N$, its Quantum Volume is defined as the exponentiation of the side length of this square, $2^N$, and the side length itself is sometimes referred to as the number of Algorithmic Qubits \cite{cha_2020_sca}. For example, a quantum device that can attain high fidelity with 4 qubits and a circuit depth of 4 has a Quantum Volume of $2^4 = 16$ and is said to have 4 Algorithmic Qubits.  

    Lastly, application-based benchmarks, which include this paper, demonstrate the abilities of a device to perform specific algorithms and tasks. A large portion of the excitement around future quantum computers is their potential for speedups in a variety of common algorithms and optimizations. As the applications for quantum devices are vast, application-based benchmarks vary dramatically. For the sake of clarity, this category of benchmarks is divided into subcategories and is presented in the following paragraphs in roughly chronological order. These subcategories are quantum chemistry, multiple task suites, visual representation, and quantum annealing.
    
    One implementation of quantum devices is the simulation of chemical reactions, which are by their nature quantum mechanical. The usefulness of this task has led to the simulation of quantum chemistry becoming a persistent benchmark for quantum computers. For example, Ref. \cite{mcc_2019_qua} utilizes simulated alkali metal hydrides to test the Variational Quantum Eigensolver on IBM and Rigetti's superconducting devices. With this approach, they were able to reach a high enough accuracy in their simulations to reproduce the results of these chemical reactions. Along a similar vein, Ref. \cite{dal_2020_ana} proposes a problem from solid-state physics, a one-dimensional Fermi-Hubbard model. In one dimension, this model has an exact solution, which makes it straightforward to test the results of the Variational Quantum Eigensolver. By layering VQE on top of itself, they managed to recreate the theoretical one-dimensional results with Google's Sycamore hardware.
    
    Later, entire suites of benchmarks were developed to test various tasks and metrics, such as QASMBench \cite{pac_2022_qas}. QASMBench is a benchmarking suite of computational problems for quantum devices based on the OpenQASM assembly language. It presents metrics in multiple tasks including linear algebra, chemistry, optimization, and cryptography. Other suites have also been developed, such as QScore \cite{a2022_qscore} and QPack \cite{mesman_2021_qpa} which both present benchmarking that centers around Max-Cut and similar optimization problems.    

    Then, building on the framework of IBM’s volumetric benchmarking, Ref. \cite{mil_2021_app} considers specific tasks for which deep (larger circuit depth) and shallow circuits (larger circuit width) would be better suited, such as state preparation and IQP circuits respectively.
    While Ref. \cite{lub_2021_app} utilizes volumetric benchmarks as a backdrop to measure a suite of quantum algorithms, such as the Quantum Fourier Transform and Grover's Search. These algorithms are tested many times at different depths and widths. Consistently, as the tasks increase in size and move farther away from the center of the quantum volume square, the fidelity drops. In this way, the relationship between a device's quantum volume and the fidelity of quantum operations is readily apparent through visually accessible figures that show when the devices begin to fail. The use of visual assessments allows for an intuitive understanding of benchmarking and has become more standard.
    For example, Ref. \cite{cor_2021_sca} tests 21 quantum devices with algorithms that were not only chosen for performance but also specifically chosen with visual accessibility in mind.
    
    Finally, benchmarks from the field of quantum annealing, which utilizes the similarity of a physical Ising model and quadratic unconstrained binary optimization (QUBO) to solve combinatorial optimization tasks, are necessary for annealing devices. Ref. \cite{tas_2022_eme} showed that specific instances of quantum annealing on DWave's \textit{Advantage} were able to produce solutions within $0.5\%$ of the best-known solution in a fraction of the time of the fastest classical algorithms. It should be noted this is not a fundamental speedup, nevertheless, order-of-magnitude differences in execution time obviously have substantial industry potential. 
     
\begin{figure*}[t]
    \centering
    \includegraphics[width=0.75\textwidth]{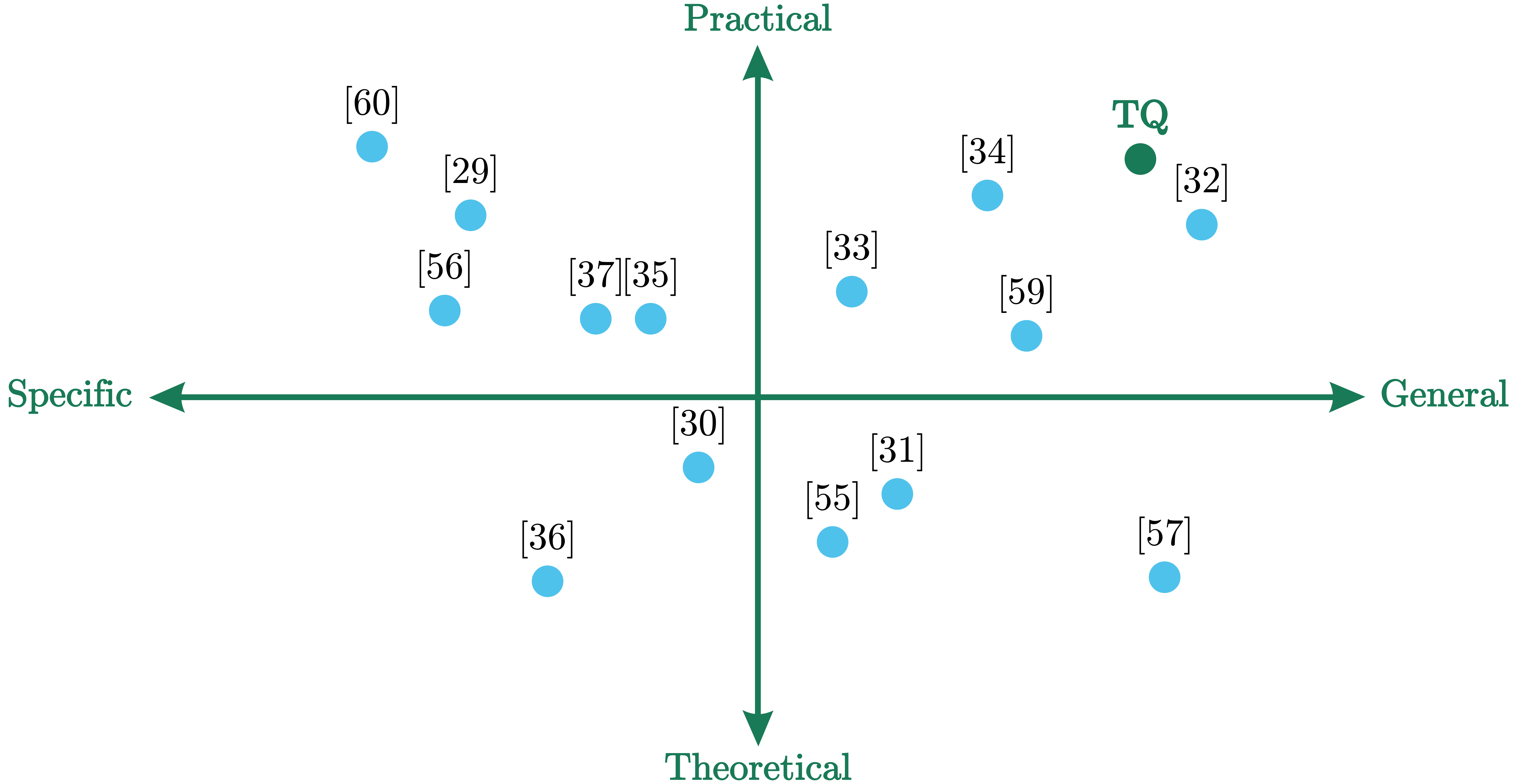}
\caption{
    Key benchmarking papers from the past 5 years plotted with specificity versus practicality. Specificity describes the requirements of hardware as well as the generality of algorithms presented by the benchmarking paper. Practicality refers to the use of metrics that apply to real-world industry problems, as opposed to theoretical problems that have been manufactured for the purposes of benchmarks themselves. It should be reiterated that all quadrants have important uses. Naturally, the placement of each paper is subjective, and others may come to different conclusions about where each benchmark belongs.}
    \label{fig:spec_v_pract}
\end{figure*}

    Application-based benchmarks span a wide field of practical industry tasks as well as theoretical problems tailored to comparing devices. Ultimately, while there have been many proposed benchmarks, they share a common theme in that they attempt to answer the degree to which a given quantum device can perform industry tasks and effectively communicate results. 
    With this in mind, Figure \ref{fig:spec_v_pract} provides an inexact categorization of these benchmarks based on the specificity of the metrics---that is, the generality of a given benchmark with respect to the hardware required and algorithms available---as well as its practicality on current noisy quantum computers. 

    This paper's testing of supervised learning on various devices serves as an application-based benchmark that has yielded preliminary results on the role of neural networks in quantum machine learning. This document tests a range of real and simulated quantum devices, comparing their time and accuracy performance on a classification task. The future role that quantum machines will play in high-level neural networks and deep learning remains to be determined, but these results demonstrate practical effects of quantum hardware as well as quantum-inspired classical hardware.


\section{Raw numerical results}
\label{sec:benchmarking-results}

In this section, the raw results of the benchmarks are provided in tabular view. Tab.~\ref{tab:data-training} demonstrates the data obtained during the training process, whereas Tab.~\ref{tab:data-inference} shows the data obtained during the inference process. The blank spaces represent two types of non-applicability: 1) the standard deviations of training (inference) runtimes were calculated up to 24 (26) qubits to avoid long execution times as well as limiting our carbon emission, and 2) as not all QPUs included enough qubits to fulfill all parts, their numbers are presented up to their highest even-qubit availability. 

Tabs.~\ref{tab:data-quantum-training-threads}, \ref{tab:data-quantum-inference-threads}, \ref{tab:data-hybrid-inference-threads}, \ref{tab:data-hybrid-training-threads} showcase the dependence of the QBN runtimes for quantum training, quantum inference, hybrid inference, and hybrid training, respectively. In all cases, it is evident that for an increasing number of threads we see a general improvement in runtimes.  However, in some cases even for low-qubit circuits it is beneficial to use more than a single thread. 

Furthermore, Tab.~\ref{tab:data-qpu-accuracies} provides the numerical results for the QPU accuracy tests, whereas the times obtained are shown in Tab.~\ref{tab:data-qpu-runtimes}.  These runtimes are largely in agreement with those in Tab.~\ref{tab:data-training}.  However, it is worth noting that these circuits are twice as deep as the ones used in runtime benchmarks as explained in Sec.~\ref{sec:accuracy}. Tab.~\ref{tab:data-large-circuit-runtime} shows the runtimes associated with running large quantum circuits. A characteristic feature of this limit is the scaling of the quantum hardware that is unavailable to any classical machine.  This hints at a decisive quantum advantage in this region with the condition that the accuracy would also become competitive. 

Finally, Tab.~\ref{tab:benchmark-qpu-prices} showcases the pricing of the various QPUs used in this work for training and inference. 
 
\begin{table*}[!ht]
\footnotesize
    \centering
    \setlength{\tabcolsep}{1pt}
    \renewcommand{\arraystretch}{1.5}
    \begin{adjustbox}{angle=90}
\begin{tabular}{|ccccccccccccccccc|}
\hline
\multicolumn{17}{|c|}{\textbf{Quantum Neural Network}} \\ \hline
\multicolumn{2}{|c|}{$n_{\text{qubits}}$} & \multicolumn{1}{c|}{\textbf{2}} & \multicolumn{1}{c|}{\textbf{4}} & \multicolumn{1}{c|}{\textbf{6}} & \multicolumn{1}{c|}{\textbf{8}} & \multicolumn{1}{c|}{\textbf{10}} & \multicolumn{1}{c|}{\textbf{12}} & \multicolumn{1}{c|}{\textbf{14}} & \multicolumn{1}{c|}{\textbf{16}} & \multicolumn{1}{c|}{\textbf{18}} & \multicolumn{1}{c|}{\textbf{20}} & \multicolumn{1}{c|}{\textbf{22}} & \multicolumn{1}{c|}{\textbf{24}} & \multicolumn{1}{c|}{\textbf{26}} & \multicolumn{1}{c|}{\textbf{28}} & \textbf{30} \\ \hline
\multicolumn{1}{|c|}{\multirow{2}{*}{\textbf{QBN}}} & \multicolumn{1}{c|}{MEAN} & \multicolumn{1}{c|}{0.0024} & \multicolumn{1}{c|}{0.0054} & \multicolumn{1}{c|}{0.0099} & \multicolumn{1}{c|}{0.0160} & \multicolumn{1}{c|}{0.0248} & \multicolumn{1}{c|}{0.0437} & \multicolumn{1}{c|}{0.0866} & \multicolumn{1}{c|}{0.1636} & \multicolumn{1}{c|}{0.3786} & \multicolumn{1}{c|}{1.2116} & \multicolumn{1}{c|}{9.8348} & \multicolumn{1}{c|}{46.9552} & \multicolumn{1}{c|}{217.3697} & \multicolumn{1}{c|}{1005.2324} & 4640.2727 \\ \cline{2-17} 
\multicolumn{1}{|c|}{} & \multicolumn{1}{c|}{STD} & \multicolumn{1}{c|}{0.0003} & \multicolumn{1}{c|}{0.0002} & \multicolumn{1}{c|}{0.0005} & \multicolumn{1}{c|}{0.0006} & \multicolumn{1}{c|}{0.0007} & \multicolumn{1}{c|}{0.0163} & \multicolumn{1}{c|}{0.0026} & \multicolumn{1}{c|}{0.0122} & \multicolumn{1}{c|}{0.0251} & \multicolumn{1}{c|}{0.0283} & \multicolumn{1}{c|}{0.0858} & \multicolumn{1}{c|}{0.4396} & \multicolumn{1}{c|}{} & \multicolumn{1}{c|}{} &  \\
\hline
\end{tabular}
    \end{adjustbox}
    \caption{Benchmarking results -- training of quantum and hybrid quantum neural networks.}
    \label{tab:data-training}
\end{table*}

\begin{table*}[!ht]
\footnotesize
    \centering
    \renewcommand{\arraystretch}{1.2}
    \begin{adjustbox}{angle=90}
\begin{tabular}{|cccccccccccccccccc|}
\hline
\multicolumn{18}{|c|}{\textbf{Quantum Neural Network}} \\ \hline
\multicolumn{3}{|c|}{\textbf{$n_{\text{qubits}}$}} & \multicolumn{1}{c|}{\textbf{2}} & \multicolumn{1}{c|}{\textbf{4}} & \multicolumn{1}{c|}{\textbf{6}} & \multicolumn{1}{c|}{\textbf{8}} & \multicolumn{1}{c|}{\textbf{10}} & \multicolumn{1}{c|}{\textbf{12}} & \multicolumn{1}{c|}{\textbf{14}} & \multicolumn{1}{c|}{\textbf{16}} & \multicolumn{1}{c|}{\textbf{18}} & \multicolumn{1}{c|}{\textbf{20}} & \multicolumn{1}{c|}{\textbf{22}} & \multicolumn{1}{c|}{\textbf{24}} & \multicolumn{1}{c|}{\textbf{26}} & \multicolumn{1}{c|}{\textbf{28}} & \textbf{30} \\ \hline
\multicolumn{1}{|c|}{\multirow{8}{*}{\textbf{Simulator}}} & \multicolumn{1}{c|}{\multirow{2}{*}{\textbf{QBN}}} & \multicolumn{1}{c|}{MEAN} & \multicolumn{1}{c|}{0.0006} & \multicolumn{1}{c|}{0.0008} & \multicolumn{1}{c|}{0.0010} & \multicolumn{1}{c|}{0.0014} & \multicolumn{1}{c|}{0.0014} & \multicolumn{1}{c|}{0.0021} & \multicolumn{1}{c|}{0.0030} & \multicolumn{1}{c|}{0.0046} & \multicolumn{1}{c|}{0.0109} & \multicolumn{1}{c|}{0.0295} & \multicolumn{1}{c|}{0.2269} & \multicolumn{1}{c|}{0.9442} & \multicolumn{1}{c|}{4.0763} & \multicolumn{1}{c|}{17.6377} & 77.0772 \\ \cline{3-18} 
\multicolumn{1}{|c|}{} & \multicolumn{1}{c|}{} & \multicolumn{1}{c|}{STD} & \multicolumn{1}{c|}{0.0000} & \multicolumn{1}{c|}{0.0002} & \multicolumn{1}{c|}{0.0003} & \multicolumn{1}{c|}{0.0001} & \multicolumn{1}{c|}{0.0004} & \multicolumn{1}{c|}{0.0008} & \multicolumn{1}{c|}{0.0006} & \multicolumn{1}{c|}{0.0007} & \multicolumn{1}{c|}{0.0009} & \multicolumn{1}{c|}{0.0018} & \multicolumn{1}{c|}{0.0135} & \multicolumn{1}{c|}{0.0271} & \multicolumn{1}{c|}{0.0520} & \multicolumn{1}{c|}{} &  \\
\hline
\end{tabular}
    \end{adjustbox}
    \caption{Benchmarking results -- inference of quantum and hybrid quantum neural networks.}
    \label{tab:data-inference}
\end{table*}

\begin{table*}[h]
\centering
\renewcommand{\arraystretch}{1.5}\centering
    \begin{adjustbox}{angle=0}
    \centering
\begin{tabular}{|ccccccccccccc|}
\hline
\multicolumn{13}{|c|}{\textbf{Pure Quantum Training}} \\ \hline
\multicolumn{3}{|c|}{\textbf{$n_{\text{qubits}}$}} & \multicolumn{1}{c|}{\textbf{2}} & \multicolumn{1}{c|}{\textbf{4}} & \multicolumn{1}{c|}{\textbf{6}} & \multicolumn{1}{c|}{\textbf{8}} & \multicolumn{1}{c|}{\textbf{10}} & \multicolumn{1}{c|}{\textbf{12}} & \multicolumn{1}{c|}{\textbf{14}} & \multicolumn{1}{c|}{\textbf{16}} & \multicolumn{1}{c|}{\textbf{18}} & \textbf{20} \\ \hline
\multicolumn{1}{|c|}{\multirow{16}{*}{\textbf{Threads}}} & \multicolumn{1}{c|}{\multirow{2}{*}{\textbf{1}}} & \multicolumn{1}{c|}{MEAN} & \multicolumn{1}{c|}{0.0024} & \multicolumn{1}{c|}{0.0054} & \multicolumn{1}{c|}{0.0099} & \multicolumn{1}{c|}{0.0160} & \multicolumn{1}{c|}{0.0248} & \multicolumn{1}{c|}{0.0437} & \multicolumn{1}{c|}{0.0964} & \multicolumn{1}{c|}{0.3243} & \multicolumn{1}{c|}{1.4042} & 6.6382 \\ \cline{3-13} 
\multicolumn{1}{|c|}{} & \multicolumn{1}{c|}{} & \multicolumn{1}{c|}{STD} & \multicolumn{1}{c|}{0.0003} & \multicolumn{1}{c|}{0.0002} & \multicolumn{1}{c|}{0.0005} & \multicolumn{1}{c|}{0.0006} & \multicolumn{1}{c|}{0.0007} & \multicolumn{1}{c|}{0.0163} & \multicolumn{1}{c|}{0.0023} & \multicolumn{1}{c|}{0.0021} & \multicolumn{1}{c|}{0.0040} & 0.0066 \\ \cline{2-13} 
\multicolumn{1}{|c|}{} & \multicolumn{1}{c|}{\multirow{2}{*}{\textbf{2}}} & \multicolumn{1}{c|}{MEAN} & \multicolumn{1}{c|}{0.0026} & \multicolumn{1}{c|}{0.0060} & \multicolumn{1}{c|}{0.0118} & \multicolumn{1}{c|}{0.0195} & \multicolumn{1}{c|}{0.0299} & \multicolumn{1}{c|}{0.0456} & \multicolumn{1}{c|}{0.1439} & \multicolumn{1}{c|}{0.2196} & \multicolumn{1}{c|}{1.1028} & 3.7856 \\
\hline
\end{tabular}
\end{adjustbox}
\caption{The performance of QMware basiq native (QBN) when training the QNN quantum network for different numbers of threads.}
\label{tab:data-quantum-training-threads}
\end{table*}

\begin{table*}
\centering
\renewcommand{\arraystretch}{1.5}\centering
    \begin{adjustbox}{angle=0}
    \centering
\begin{tabular}{|cccllllllllll|}
\hline
\multicolumn{13}{|c|}{\textbf{Pure Quantum Forward}} \\ \hline
\multicolumn{3}{|c|}{\textbf{$n_{\text{qubits}}$}} & \multicolumn{1}{c|}{\textbf{2}} & \multicolumn{1}{c|}{\textbf{4}} & \multicolumn{1}{c|}{\textbf{6}} & \multicolumn{1}{c|}{\textbf{8}} & \multicolumn{1}{c|}{\textbf{10}} & \multicolumn{1}{c|}{\textbf{12}} & \multicolumn{1}{c|}{\textbf{14}} & \multicolumn{1}{c|}{\textbf{16}} & \multicolumn{1}{c|}{\textbf{18}} & \multicolumn{1}{c|}{\textbf{20}} \\ \hline
\multicolumn{1}{|c|}{\multirow{16}{*}{\textbf{Threads}}} & \multicolumn{1}{c|}{\multirow{2}{*}{\textbf{1}}} & \multicolumn{1}{c|}{MEAN} & \multicolumn{1}{l|}{0.0008} & \multicolumn{1}{l|}{0.0007} & \multicolumn{1}{l|}{0.0009} & \multicolumn{1}{l|}{0.0011} & \multicolumn{1}{l|}{0.0015} & \multicolumn{1}{l|}{0.0020} & \multicolumn{1}{l|}{0.0034} & \multicolumn{1}{l|}{0.0099} & \multicolumn{1}{l|}{0.0400} & 0.1687 \\ \cline{3-13} 
\multicolumn{1}{|c|}{} & \multicolumn{1}{c|}{} & \multicolumn{1}{c|}{STD} & \multicolumn{1}{l|}{0.0001} & \multicolumn{1}{l|}{0.0001} & \multicolumn{1}{l|}{0.0002} & \multicolumn{1}{l|}{0.0003} & \multicolumn{1}{l|}{0.0005} & \multicolumn{1}{l|}{0.0006} & \multicolumn{1}{l|}{0.0008} & \multicolumn{1}{l|}{0.0008} & \multicolumn{1}{l|}{0.0148} & 0.0087 \\ \cline{2-13} 
\multicolumn{1}{|c|}{} & \multicolumn{1}{c|}{\multirow{2}{*}{\textbf{2}}} & \multicolumn{1}{c|}{MEAN} & \multicolumn{1}{l|}{0.0007} & \multicolumn{1}{l|}{0.0008} & \multicolumn{1}{l|}{0.0010} & \multicolumn{1}{l|}{0.0012} & \multicolumn{1}{l|}{0.0030} & \multicolumn{1}{l|}{0.0020} & \multicolumn{1}{l|}{0.0030} & \multicolumn{1}{l|}{0.0068} & \multicolumn{1}{l|}{0.0236} & 0.0935 \\
\hline
\end{tabular}\end{adjustbox}
\caption{The performance of QMware basiq native (QBN) when performing inference using the QNN quantum network for different numbers of threads.}
\label{tab:data-quantum-inference-threads}
\end{table*}

\begin{table*}
\centering
\renewcommand{\arraystretch}{1.5}\centering
    \begin{adjustbox}{angle=0}
    \centering
\begin{tabular}{|cccllllllllll|}
\hline
\multicolumn{13}{|c|}{\textbf{Hybrid Quantum Forward}} \\ \hline
\multicolumn{3}{|c|}{\textbf{$n_{\text{qubits}}$}} & \multicolumn{1}{c|}{\textbf{2}} & \multicolumn{1}{c|}{\textbf{4}} & \multicolumn{1}{c|}{\textbf{6}} & \multicolumn{1}{c|}{\textbf{8}} & \multicolumn{1}{c|}{\textbf{10}} & \multicolumn{1}{c|}{\textbf{12}} & \multicolumn{1}{c|}{\textbf{14}} & \multicolumn{1}{c|}{\textbf{16}} & \multicolumn{1}{c|}{\textbf{18}} & \multicolumn{1}{c|}{\textbf{20}} \\ \hline
\multicolumn{1}{|c|}{\multirow{16}{*}{\textbf{Threads}}} & \multicolumn{1}{c|}{\multirow{2}{*}{\textbf{1}}} & \multicolumn{1}{c|}{MEAN} & \multicolumn{1}{l|}{0.0008} & \multicolumn{1}{l|}{0.0008} & \multicolumn{1}{l|}{0.0010} & \multicolumn{1}{l|}{0.0028} & \multicolumn{1}{l|}{0.0014} & \multicolumn{1}{l|}{0.0021} & \multicolumn{1}{l|}{0.0036} & \multicolumn{1}{l|}{0.0104} & \multicolumn{1}{l|}{0.0419} & 0.1680 \\ \cline{3-13} 
\multicolumn{1}{|c|}{} & \multicolumn{1}{c|}{} & \multicolumn{1}{c|}{STD} & \multicolumn{1}{l|}{0.0001} & \multicolumn{1}{l|}{0.0002} & \multicolumn{1}{l|}{0.0003} & \multicolumn{1}{l|}{0.0159} & \multicolumn{1}{l|}{0.0004} & \multicolumn{1}{l|}{0.0008} & \multicolumn{1}{l|}{0.0008} & \multicolumn{1}{l|}{0.0013} & \multicolumn{1}{l|}{0.0168} & 0.0105 \\ \cline{2-13} 
\multicolumn{1}{|c|}{} & \multicolumn{1}{c|}{\multirow{2}{*}{\textbf{2}}} & \multicolumn{1}{c|}{MEAN} & \multicolumn{1}{l|}{0.0007} & \multicolumn{1}{l|}{0.0010} & \multicolumn{1}{l|}{0.0011} & \multicolumn{1}{l|}{0.0014} & \multicolumn{1}{l|}{0.0016} & \multicolumn{1}{l|}{0.0022} & \multicolumn{1}{l|}{0.0331} & \multicolumn{1}{l|}{0.0071} & \multicolumn{1}{l|}{0.0242} & 0.0932 \\
\hline
\end{tabular}

\end{adjustbox}
\caption{The performance of QMware basiq native (QBN) when performing inference using the HQNN for different numbers of threads.}
\label{tab:data-hybrid-inference-threads}
\end{table*}

\begin{table*}
\centering
\renewcommand{\arraystretch}{1.5}\centering
    \begin{adjustbox}{angle=0}
    \centering
\begin{tabular}{|ccccccccccccc|}
\hline
\multicolumn{13}{|c|}{\textbf{Hybrid Quantum Training}} \\ \hline
\multicolumn{3}{|c|}{\textbf{$n_{\text{qubits}}$}} &
  \multicolumn{1}{c|}{\textbf{2}} &
  \multicolumn{1}{c|}{\textbf{4}} &
  \multicolumn{1}{c|}{\textbf{6}} &
  \multicolumn{1}{c|}{\textbf{8}} &
  \multicolumn{1}{c|}{\textbf{10}} &
  \multicolumn{1}{c|}{\textbf{12}} &
  \multicolumn{1}{c|}{\textbf{14}} &
  \multicolumn{1}{c|}{\textbf{16}} &
  \multicolumn{1}{c|}{\textbf{18}} &
  \textbf{20} \\ \hline
\multicolumn{1}{|c|}{\multirow{16}{*}{\textbf{Threads}}} &
  \multicolumn{1}{c|}{\multirow{2}{*}{\textbf{1}}} &
  \multicolumn{1}{c|}{MEAN} &
  \multicolumn{1}{c|}{0.0008} &
  \multicolumn{1}{c|}{0.0007} &
  \multicolumn{1}{c|}{0.0009} &
  \multicolumn{1}{c|}{0.0011} &
  \multicolumn{1}{c|}{0.0015} &
  \multicolumn{1}{c|}{0.0020} &
  \multicolumn{1}{c|}{0.0034} &
  \multicolumn{1}{c|}{0.0099} &
  \multicolumn{1}{c|}{0.0400} &
  0.1687 \\ \cline{3-13} 
\multicolumn{1}{|c|}{} &
  \multicolumn{1}{c|}{} &
  \multicolumn{1}{c|}{STD} &
  \multicolumn{1}{c|}{0.0001} &
  \multicolumn{1}{c|}{0.0001} &
  \multicolumn{1}{c|}{0.0002} &
  \multicolumn{1}{c|}{0.0003} &
  \multicolumn{1}{c|}{0.0005} &
  \multicolumn{1}{c|}{0.0006} &
  \multicolumn{1}{c|}{0.0008} &
  \multicolumn{1}{c|}{0.0008} &
  \multicolumn{1}{c|}{0.0148} &
  0.0087 \\ \cline{2-13} 
\multicolumn{1}{|c|}{} &
  \multicolumn{1}{c|}{\multirow{2}{*}{\textbf{2}}} &
  \multicolumn{1}{c|}{MEAN} &
  \multicolumn{1}{c|}{0.0007} &
  \multicolumn{1}{c|}{0.0008} &
  \multicolumn{1}{c|}{0.0010} &
  \multicolumn{1}{c|}{0.0012} &
  \multicolumn{1}{c|}{0.0030} &
  \multicolumn{1}{c|}{0.0020} &
  \multicolumn{1}{c|}{0.0030} &
  \multicolumn{1}{c|}{0.0068} &
  \multicolumn{1}{c|}{0.0236} &
  0.0935 \\
\hline
\end{tabular}
\end{adjustbox}
\caption{The performance of QMware basiq native (QBN) when training the hybrid quantum network for different numbers of threads.}
\label{tab:data-hybrid-training-threads}
\end{table*}

\begin{table*}
\begin{tabular}{|clcccccccccc|}
\hline
\multicolumn{12}{|c|}{\textbf{Accuracy, \%}} \\ \hline
\multicolumn{3}{|c|}{$n_{\text{qubits}}$} & \multicolumn{1}{c|}{2} & \multicolumn{1}{c|}{4} & \multicolumn{1}{c|}{6} & \multicolumn{1}{c|}{8} & \multicolumn{1}{c|}{10} & \multicolumn{1}{c|}{12} & \multicolumn{1}{c|}{14} & \multicolumn{1}{c|}{16} & 18 \\ \hline
\multicolumn{2}{|c|}{\multirow{2}{*}{IonQ Harmony}} & \multicolumn{1}{c|}{MEAN} & \multicolumn{1}{c|}{99.76} & \multicolumn{1}{c|}{99.73} & \multicolumn{1}{c|}{99.52} & \multicolumn{1}{c|}{99.19} & \multicolumn{1}{c|}{99.12} & \multicolumn{1}{c|}{} & \multicolumn{1}{c|}{} & \multicolumn{1}{c|}{} &  \\ \cline{3-12} 
\multicolumn{2}{|c|}{} & \multicolumn{1}{c|}{STD} & \multicolumn{1}{c|}{0.20} & \multicolumn{1}{c|}{0.17} & \multicolumn{1}{c|}{0.12} & \multicolumn{1}{c|}{0.27} & \multicolumn{1}{c|}{0.30} & \multicolumn{1}{c|}{} & \multicolumn{1}{c|}{} & \multicolumn{1}{c|}{} &  \\ \hline
\multicolumn{2}{|c|}{\multirow{2}{*}{Rigetti Aspen-M2}} & \multicolumn{1}{c|}{MEAN} & \multicolumn{1}{c|}{88.77} & \multicolumn{1}{c|}{5.47} & \multicolumn{1}{c|}{0.67} & \multicolumn{1}{c|}{0.15} & \multicolumn{1}{c|}{0.04} & \multicolumn{1}{c|}{0.02} & \multicolumn{1}{c|}{0.00} & \multicolumn{1}{c|}{0.00} & 0.00 \\ \cline{3-12} 
\multicolumn{2}{|c|}{} & \multicolumn{1}{c|}{STD} & \multicolumn{1}{c|}{6.81} & \multicolumn{1}{c|}{0.75} & \multicolumn{1}{c|}{2.65} & \multicolumn{1}{c|}{0.12} & \multicolumn{1}{c|}{0.07} & \multicolumn{1}{c|}{0.04} & \multicolumn{1}{c|}{0.00} & \multicolumn{1}{c|}{0.00} & 0.00 \\ \hline
\multicolumn{2}{|c|}{\multirow{2}{*}{OQC Lucy}} & \multicolumn{1}{c|}{MEAN} & \multicolumn{1}{c|}{93.76} & \multicolumn{1}{c|}{67.87} & \multicolumn{1}{c|}{63.66} & \multicolumn{1}{c|}{57.64} & \multicolumn{1}{c|}{} & \multicolumn{1}{c|}{} & \multicolumn{1}{c|}{} & \multicolumn{1}{c|}{} &  \\
\hline
\end{tabular}
\caption{QPU accuracies}
\label{tab:data-qpu-accuracies}
\end{table*}

\begin{table*}
\begin{tabular}{|clcccccccccc|}
\hline
\multicolumn{12}{|c|}{\textbf{Times, seconds}} \\ \hline
\multicolumn{3}{|c|}{$n_{\text{qubits}}$} & \multicolumn{1}{c|}{2} & \multicolumn{1}{c|}{4} & \multicolumn{1}{c|}{6} & \multicolumn{1}{c|}{8} & \multicolumn{1}{c|}{10} & \multicolumn{1}{c|}{12} & \multicolumn{1}{c|}{14} & \multicolumn{1}{c|}{16} & 18 \\ \hline
\multicolumn{2}{|c|}{\multirow{2}{*}{IonQ Harmony}} & \multicolumn{1}{c|}{MEAN} & \multicolumn{1}{c|}{350.2784} & \multicolumn{1}{c|}{80.0766} & \multicolumn{1}{c|}{339.8358} & \multicolumn{1}{c|}{17.9963} & \multicolumn{1}{c|}{73.2473} & \multicolumn{1}{c|}{} & \multicolumn{1}{c|}{} & \multicolumn{1}{c|}{} &  \\ \cline{3-12} 
\multicolumn{2}{|c|}{} & \multicolumn{1}{c|}{STD} & \multicolumn{1}{c|}{501.8712} & \multicolumn{1}{c|}{59.0565} & \multicolumn{1}{c|}{630.4188} & \multicolumn{1}{c|}{7.0536} & \multicolumn{1}{c|}{102.1677} & \multicolumn{1}{c|}{} & \multicolumn{1}{c|}{} & \multicolumn{1}{c|}{} &  \\ \hline
\multicolumn{2}{|c|}{\multirow{2}{*}{Rigetti Aspen-M2}} & \multicolumn{1}{c|}{MEAN} & \multicolumn{1}{c|}{58.7262} & \multicolumn{1}{c|}{52.1269} & \multicolumn{1}{c|}{60.1630} & \multicolumn{1}{c|}{64.3308} & \multicolumn{1}{c|}{58.6161} & \multicolumn{1}{c|}{85.4907} & \multicolumn{1}{c|}{85.7971} & \multicolumn{1}{c|}{81.8377} & 99.0498 \\ \cline{3-12} 
\multicolumn{2}{|c|}{} & \multicolumn{1}{c|}{STD} & \multicolumn{1}{c|}{35.1338} & \multicolumn{1}{c|}{21.8146} & \multicolumn{1}{c|}{17.8216} & \multicolumn{1}{c|}{16.9935} & \multicolumn{1}{c|}{20.6759} & \multicolumn{1}{c|}{26.5828} & \multicolumn{1}{c|}{22.4378} & \multicolumn{1}{c|}{33.9639} & 19.8638 \\ \hline
\multicolumn{2}{|c|}{\multirow{2}{*}{OQC Lucy}} & \multicolumn{1}{c|}{MEAN} & \multicolumn{1}{c|}{3.4535} & \multicolumn{1}{c|}{3.3095} & \multicolumn{1}{c|}{3.3682} & \multicolumn{1}{c|}{3.4450} & \multicolumn{1}{c|}{} & \multicolumn{1}{c|}{} & \multicolumn{1}{c|}{} & \multicolumn{1}{c|}{} &  \\
\hline
\end{tabular}
\caption{QPUs runtimes for the accuracy test}
\label{tab:data-qpu-runtimes}
\end{table*}

\begin{table*}
\begin{adjustbox}{angle=0}
\renewcommand{\arraystretch}{1.2}\centering
\begin{tabular}{|ccc|c|c|c|c|c|}
\hline
\multicolumn{3}{|c|}{\textbf{$n_{\text{qubits}}$}} & \textbf{32} & \textbf{34} & \textbf{36} & \textbf{38} & \textbf{40} \\ \hline
\multicolumn{1}{|c|}{\multirow{4}{*}{\textbf{QNN}}} & \multicolumn{2}{c|}{\textbf{QBN}} & 304.1323 & 1343.5551 & 5952.5783 & 28893.5688 & 48970.1862 \\ \cline{2-8} 
\multicolumn{1}{|c|}{} & \multicolumn{2}{c|}{\textbf{ASV}} & 237.3554 & 986.5651 &  &  &  \\ \cline{2-8} 
\multicolumn{1}{|c|}{} & \multicolumn{1}{c|}{\multirow{2}{*}{\textbf{Rigetti Aspen-M2}}} & MEAN & 51.9887 & 52.1990 & 59.0934 & 60.9397 & 73.9763 \\ \cline{3-8} 
\multicolumn{1}{|c|}{} & \multicolumn{1}{c|}{} & STD & 8.9592 & 7.9419 & 11.5971 & 7.4355 & 11.4524 \\ \hline
\multicolumn{1}{|c|}{\multirow{4}{*}{\textbf{HQNN}}} & \multicolumn{2}{c|}{\textbf{QBN}} & 270.6306 & 1244.3166 & 3450.6118 & 18630.4505 & 47273.2937 \\ \cline{2-8} 
\multicolumn{1}{|c|}{} & \multicolumn{2}{c|}{\textbf{ASV}} & 323.9802 & 1414.4067 &  &  &  \\ \cline{2-8} 
\multicolumn{1}{|c|}{} & \multicolumn{1}{c|}{\multirow{2}{*}{\textbf{Rigetti Aspen-M2}}} & MEAN & 49.9818 & 52.4805 & 61.9770 & 64.4956 & 67.4168 \\ \cline{3-8} 
\multicolumn{1}{|c|}{} & \multicolumn{1}{c|}{} & STD & 6.9296 & 7.0688 & 17.2564 & 7.6582 & 7.9843 \\ \hline
\end{tabular}
\end{adjustbox}
\caption{Inference runtimes beyond 30 qubits}
\label{tab:data-large-circuit-runtime}
\end{table*}

\begin{table*}
\centering
\renewcommand{\arraystretch}{1.5}\centering
    \begin{adjustbox}{angle=90}
    \centering
\begin{tabular}{|cccccccccccccccc|}
\hline
\multicolumn{16}{|c|}{\textbf{Prices, inference in US dollars and training in thousands of US dollars}} \\ \hline
\multicolumn{3}{|c|}{\textbf{$n_{\text{qubits}}$}} & \multicolumn{1}{c|}{\textbf{2}} & \multicolumn{1}{c|}{\textbf{4}} & \multicolumn{1}{c|}{\textbf{6}} & \multicolumn{1}{c|}{\textbf{8}} & \multicolumn{1}{c|}{\textbf{10}} & \multicolumn{1}{c|}{\textbf{12}} & \multicolumn{1}{c|}{\textbf{14}} & \multicolumn{1}{c|}{\textbf{16}} & \multicolumn{1}{c|}{\textbf{18}} & \multicolumn{1}{c|}{\textbf{20}} & \multicolumn{1}{c|}{\textbf{22}} & \multicolumn{1}{c|}{\textbf{24}} & \textbf{26} \\ \hline
\multicolumn{1}{|c|}{\multirow{2}{*}{\textbf{IonQ Harmony}}} & \multicolumn{2}{c|}{Inference} & \multicolumn{1}{c|}{10.30} & \multicolumn{1}{c|}{10.30} & \multicolumn{1}{c|}{10.30} & \multicolumn{1}{c|}{10.30} & \multicolumn{1}{c|}{10.30} & \multicolumn{1}{c|}{} & \multicolumn{1}{c|}{} & \multicolumn{1}{c|}{} & \multicolumn{1}{c|}{} & \multicolumn{1}{c|}{} & \multicolumn{1}{c|}{} & \multicolumn{1}{c|}{} &  \\ \cline{2-16} 
\multicolumn{1}{|c|}{} & \multicolumn{2}{c|}{Training} & \multicolumn{1}{c|}{515.00} & \multicolumn{1}{c|}{927.00} & \multicolumn{1}{c|}{1339.00} & \multicolumn{1}{c|}{1751.00} & \multicolumn{1}{c|}{2163.00} & \multicolumn{1}{c|}{} & \multicolumn{1}{c|}{} & \multicolumn{1}{c|}{} & \multicolumn{1}{c|}{} & \multicolumn{1}{c|}{} & \multicolumn{1}{c|}{} & \multicolumn{1}{c|}{} &  \\ \hline
\multicolumn{1}{|c|}{\multirow{2}{*}{\textbf{Rigetti Aspen-M2}}} & \multicolumn{2}{c|}{Inference} & \multicolumn{1}{c|}{0.34} & \multicolumn{1}{c|}{0.34} & \multicolumn{1}{c|}{0.34} & \multicolumn{1}{c|}{0.34} & \multicolumn{1}{c|}{0.34} & \multicolumn{1}{c|}{0.34} & \multicolumn{1}{c|}{0.34} & \multicolumn{1}{c|}{0.34} & \multicolumn{1}{c|}{0.34} & \multicolumn{1}{c|}{0.34} & \multicolumn{1}{c|}{0.34} & \multicolumn{1}{c|}{0.34} & 0.34 \\
\hline
\end{tabular}
\end{adjustbox}
\caption{Cost of executing inference and training runs on QPUs.}
\label{tab:benchmark-qpu-prices}
\end{table*}

\end{document}